%
\font\fourteenrm=cmr10 scaled \magstep2
\font\fourteeni=cmmi10 scaled \magstep2
\font\fourteenbf=cmbx10 scaled \magstep2
\font\fourteenit=cmti10 scaled \magstep2
\font\fourteensy=cmsy10 scaled \magstep2
\font\large=cmbx10 scaled \magstep1
%

%
%

%
%
\font\bdi=cmmib10
%
%
\def\bi#1{\hbox{\bdi #1\/}}
%
%
\font\eightrm=cmr8
\font\eighti=cmmi8
\font\eightbf=cmbx8
\font\eightit=cmti8

\font\eightsy=cmsy8
\font\sixrm=cmr6
\font\sixi=cmmi6
\font\sixsy=cmsy6

\def\tenpoint{\def\rm{\fam0\tenrm}%
  \textfont0=\tenrm \scriptfont0=\sevenrm
                      \scriptscriptfont0=\fiverm
  \textfont1=\teni  \scriptfont1=\seveni
                      \scriptscriptfont1=\fivei
  \textfont2=\tensy \scriptfont2=\sevensy
                      \scriptscriptfont2=\fivesy
  \textfont3=\tenex   \scriptfont3=\tenex
                      \scriptscriptfont3=\tenex
  \textfont\itfam=\tenit  \def\it{\fam\itfam\tenit}%
  \textfont\slfam=\tensl  \def\sl{\fam\slfam\tensl}%
  \textfont\bffam=\tenbf  \scriptfont\bffam=\sevenbf
                            \scriptscriptfont\bffam=\fivebf
                            \def\bf{\fam\bffam\tenbf}%
  \normalbaselineskip=22 truept
  \setbox\strutbox=\hbox{\vrule height14pt depth6pt
width0pt}%
  \let\sc=\eightrm \normalbaselines\rm}
\def\eightpoint{\def\rm{\fam0\eightrm}%
  \textfont0=\eightrm \scriptfont0=\sixrm
                      \scriptscriptfont0=\fiverm
  \textfont1=\eighti  \scriptfont1=\sixi
                      \scriptscriptfont1=\fivei
  \textfont2=\eightsy \scriptfont2=\sixsy
                      \scriptscriptfont2=\fivesy
  \textfont3=\tenex   \scriptfont3=\tenex
                      \scriptscriptfont3=\tenex
  \textfont\itfam=\eightit  \def\it{\fam\itfam\eightit}%
  \textfont\bffam=\eightbf  \def\bf{\fam\bffam\eightbf}%
  \normalbaselineskip=16 truept
  \setbox\strutbox=\hbox{\vrule height11pt depth5pt width0pt}}
\def\fourteenpoint{\def\rm{\fam0\fourteenrm}%
  \textfont0=\fourteenrm \scriptfont0=\tenrm
                      \scriptscriptfont0=\eightrm
  \textfont1=\fourteeni  \scriptfont1=\teni
                      \scriptscriptfont1=\eighti
  \textfont2=\fourteensy \scriptfont2=\tensy
                      \scriptscriptfont2=\eightsy
  \textfont3=\tenex   \scriptfont3=\tenex
                      \scriptscriptfont3=\tenex
  \textfont\itfam=\fourteenit  \def\it{\fam\itfam\fourteenit}%
  \textfont\bffam=\fourteenbf  \scriptfont\bffam=\tenbf
                             \scriptscriptfont\bffam=\eightbf
                             \def\bf{\fam\bffam\fourteenbf}%
  \setbox\strutbox=\hbox{\vrule height17pt depth7pt width0pt}%
  \let\sc=\tenrm \normalbaselines\rm}
  \normalbaselineskip=16 truept
  \baselineskip=16 truept

%
\newcount\secno      
\newcount\subno      
\newcount\subsubno   

%
\def\title#1
   {\vglue1truein
   {\baselineskip=24 truept
    \pretolerance=10000
    \raggedright
    \noindent \fourteenpoint\bf #1\par}
    \vskip1truein minus36pt}
%
%
\def\author#1
  {{\pretolerance=10000
    \raggedright
    \noindent {\large #1}\par}}
%
%
\def\address#1
   {\bigskip
    \noindent \rm #1\par}
%
%
\def\shorttitle#1
   {\vfill
    \noindent \rm Short title: {\sl #1}\par
    \medskip}
%
%
\def\pacs#1
   {\noindent \rm PACS number(s): #1\par
    \medskip}
%
%
\def\jnl#1
   {\noindent \rm Submitted to: {\sl #1}\par
    \medskip}
%
%

%
%

%
\def\section#1
   {\vskip0pt plus.1\vsize\penalty-250
    \vskip0pt plus-.1\vsize\vskip24pt plus12pt minus6pt
    \subno=0 \subsubno=0
    \global\advance\secno by 1
    \noindent {\bf \the\secno. #1\par}
    \bigskip
    \noindent}
%
%
\def\ack
   {\vskip-\lastskip
    \vskip36pt plus12pt minus12pt
    \bigbreak
    \noindent{\bf Acknowledgments\par}
    \nobreak
    \bigskip
    \noindent}
%
%
\def\references
     {\vfill\eject
     {\noindent \bf References\par}
      \parindent=0pt
      \bigskip}
%
%
\def\refjl#1#2#3#4
   {\hangindent=16pt
    \hangafter=1
    \rm #1
    {\sl #2
    \bf #3},
    #4\par}
%
%
\def\refbk#1#2#3
   {\hangindent=16pt
    \hangafter=1
    \rm #1
   {\frenchspacing\sl #2}
    #3\par}
%
%
\def\dash{---{}---}
%
%
\def\i{\ifmmode{\rm i}\else\char"10\fi}
%
%
\def\CQG{Classical Quantum Grav.}
\def\CJP{Can. J. Phys.}
\def\JMP{J. Math. Phys.}
\def\PL{Phys. Lett.}
\def\PR{Phys. Rev.}
\def\PRL{Phys. Rev. Lett.}

\fontdimen16\tensy=2.8pt
\fontdimen17\tensy=2.8pt

\font\boldgr=cmmib10
\def\bgr#1{\hbox{$\textfont1=\boldgr #1$}}

\def \tx #1{\textstyle{#1}}
\def \sym #1{{\sevenrm (}#1{\sevenrm )}}
\def \asym #1{{\sevenrm [}#1{\sevenrm ]}}

\def\bhat#1{\hbox{\bf\^{\bdi #1}}}
\def\ie{\hbox{\it i.e.,\ }}

\def\etal{\hbox{\it et al\ }}

\def\dn#1{\lower 2 pt \hbox{$\scriptstyle\rm {#1}$}}
\def\dnt#1{\lower 2 pt \hbox{$ {#1}$}}

\def\Adot {\dot A}
\def\psidot {\dot \psi}
\def\epspar {\epsilon_{\parallel}}
\def\epsperp {\epsilon_{\perp}}
\def\mupar {\mu_{\parallel}}
\def\muperp {\mu_{\perp}}

\title{Anisotropic Homogeneous Cosmology in the Nonsymmetric Theory
of Gravitation}

\author{Pierre Savaria}

\address{Department of Physics, University of Toronto, Toronto,
Canada M5S 1A7}

\vskip 1 true cm

\jnl{Classical and Quantum Gravity}

\noindent{October 1997}

\vskip 2 true cm

\noindent{\bf Abstract}

     
Solutions of the field equations of the Nonsymmetric Gravitational
Theory with ${\sl g}_{\asym{0i}} = 0$ are obtained for the
homogeneous, plane-symmetric, time-dependent case, both in vacuum
and in the presence of a perfect fluid. Cosmological consequences
include a dependence of the speed of light on its polarisation,
as in a birefringent crystal.

\par\vfill\eject

\section{Introduction}

As an alternative to Einstein's theory of gravitation (GR),
the Nonsymmetric Theory of Gravitation
(NGT) possesses a richer structure with the potential for
significantly different physical predictions. This structure, first 
proposed by Einstein and Straus (1945, 1946) in an attempt to unify gravitation
and electromagnetism, is characterised by a nonsymmetric
fundamental tensor ${\bi g}$. The tangent bundle of spacetime is
endowed with a hypercomplex algebraic structure (Crumeyrolle 1967, Kunstatter
\etal 1983, Mann 1984, 1989, Moffat 1984) compatible with ${\bi g}$
and with the connections. In terms of components, one has
${\sl g}_{\mu\nu} = {\sl g}_{\sym{\mu\nu}} + e\,{\sl g}_{\asym{\mu\nu}}$,
where $e^2 = +1$, and round and square brackets stand for symmetrisation
and antisymmetrisation, respectively.

Although this attempt at unification was not successful, Moffat
(1979) suggested that the extended structure could be interpreted as
a generalised theory of gravitation. For several years thereafter, 
the physical content of the theory was thought (Moffat 1991) to arise from 
the existence of a new conserved charge which entered in the time-space
components (${\sl g}_{\asym{i0}}$) of ${\bi g}$, as well as in 
its diagonal ones (line element), 
but with the space-space components (${\sl g}_{\asym{ij}}$)
being put equal to zero. After it was pointed out (Damour \etal 1992, 1993)
that with this choice, the theory, unprotected by gauge invariance,
would allow coupled negative-energy modes to propagate to infinity,
a version with ${\sl g}_{\asym{ij}}\neq0$ was proposed
(Cornish and Moffat 1994, L\'egar\'e and Moffat 1995, 
Moffat 1995a, 1995b, 1995c) 
which avoids this defect, as well as other stability problems 
(Clayton 1996, 1997), as long as the ${\sl g}_{\asym{i0}}$ can be put equal 
to zero. Claims (Moffat and Sokolov 1995, Moffat 1995c) 
and counterclaims (Burko and Ori 1995) have been made 
since then about the absence of apparent event horizons 
under dynamical conditions in the spherically-symmetric case.
Unfortunately, only the static vacuum spherically-symmetric 
solution has been obtained analytically (Wyman 1950)
for ${\sl g}_{\asym{i0}}  = 0$, and it is difficult to infer from it 
the behaviour of time-dependent solutions.  

This paper presents the first such analytic dynamical solution, but
instead of the spherically-symmetric case, we address the 
more tractable homogeneous plane-symmetric case which may 
have applications to cosmology. Section 2 sets the
relevant field equations to be solved. Very little theoretical background 
is provided as this is easily accessible in the existing literature
(see for instance Moffat 1995b). Section 3 presents the solution to
the vacuum case, and Section 4 the solution to the case with
matter. Section 5 contains a few comments about the matter solution,
in particular expressions containing NGT corrections for the GR
prediction for the rate
of expansion of the universe. In Section 6 we derive one major
consequence of the direct coupling of NGT to electromagnetic fields:
the universe can behave somewhat like a birefringent crystal
so that the polarisation vector of light is rotated as it travels
over cosmic distances. The paper ends with a few concluding remarks
in Section 7.

\section{Field Equations}

In this paper, boldface symbols denote tensors, while a tilde above 
a tensor makes it a density. We work in units where  $G = c = 1$. 
Finally, $\dot X \equiv X_{,t} \equiv \partial X/\partial t$.

In the homogeneous, plane-symmetric case, we use the fundamental tensor:
$$
{\bi g} = e^\nu dt\otimes dt - e^{\lambda}dx\otimes dx - \beta(dy\otimes dy +
dz\otimes dz) + f(dy \wedge dz),\eqno(2.1)
$$
where, without loss of generality, the $yz$ plane has been chosen as the
plane of symmetry, and $\nu$, $\lambda$, $\beta$ and $f$ are functions
of time.

Denoting by ${\sl g}^{\mu\nu}$
the contravariant components of ${\bi g}$, the {\it homogeneous,
plane-symmetric} connection components obey
$$
{\tilde{\sl g}}^{\mu\nu}{}_{\/,\lambda} + 
\tilde{\sl g}^{\rho\nu}{\mit\Gamma}^\mu_{\rho\lambda} + 
\tilde{\sl g}^{\mu\rho}{\mit\Gamma}^\nu_{\lambda\rho} - 
\tilde{\sl g}^{\mu\nu}{\mit\Gamma}^\rho_{\sym{\lambda\rho}} = 0\eqno(2.2)
$$

Define the Einstein tensor, ${\bi G}\equiv
{\bi R} - {1\over2}{\bi g}\,{\hbox{Tr}}{\bi R}$, where the covariant
components of ${\bi R}$ are given by 
$$
R^{\phantom\nu}_{\mu\nu} = {\mit\Gamma}^{\alpha}_{\mu\nu,\alpha} 
        - \tx{1\over2}({\mit\Gamma}^{\alpha}_{\mu\alpha,\nu}
        + {\mit\Gamma}^{\alpha}_{\nu\alpha,\mu})
        - {\mit\Gamma}^{\alpha}_{\mu\beta}{\mit\Gamma}^{\beta}_{\alpha\nu} 
        + \,{\mit\Gamma}^{\alpha}_{\mu\nu}{\mit\Gamma}^{\beta}_{\alpha\beta}
\eqno(2.3)
$$

Then, the other homogeneous, plane-symmetric field equations reduce to
$$
{\bi G}\ =\ 8\pi{\bi T} \eqno(2.4) 
$$
where a possible contribution from a cosmological constant has been ignored.

We assume matter to consist of a perfect fluid made of particles moving
with four-velocity $\bi u$. The energy-momentum tensor takes the form
(Vincent 1985):
$$
{\bi T}\ = (\rho + p){\bi u}\otimes{\bi u} - p{\bi g},\eqno(2.5)
$$
with $\rho$ the internal energy density and $p$ the pressure.
We choose a comoving frame, where $u^i = 0$ $(i= x,y,z)$, and
$u^t = 1$. Since only three of the four equations (2.4) will be
independent, one of the four functions in (2.1) is arbitrary;
it makes sense to take $\nu =0$. The time $t$ is interpreted as
the proper time of the particles in the fluid.

Now, introduce the
parametrisation $\beta = R^2\cos\psi$, and $f = R^2\sin\psi$,
as well as the definition $A = \log R^2$. 
The nonvanishing connection components which solve (2.2) read

$$
\eqalignno{
{\mit\Gamma}^t_{xx}\ &= \tx{1\over2}\dot\lambda e^{\lambda},
\qquad\qquad{\mit\Gamma}^t_{yy} = {\mit\Gamma}^t_{zz} = 
\tx{1\over2}(\beta\Adot - f\psidot)\cr
{\mit\Gamma}^x_{\sym{tx}} &= \tx{1\over2}\dot\lambda,\qquad\qquad
{\mit\Gamma}^t_{\asym{yz}}\ = -\tx{1\over2}(f\Adot + \beta\psidot)\cr
{\mit\Gamma}^z_{\sym{tz}} &= \tx{1\over2}\dot A,\qquad\qquad
{\mit\Gamma}^y_{\asym{tz}} = -{\mit\Gamma}^z_{\asym{ty}}= 
\tx{1\over2}\psidot&(2.6)\cr}
$$

With the help of (2.3) and (2.6), (2.4) can now be written down in terms
of the components of ${\bi g}$. Indeed, the nonzero mixed components 
of {\bi G} are

$$
{G_t}^t\ =  \tx{1\over4}(\Adot^2 - \psidot^2) +
 \tx{1\over2}\Adot\dot\lambda = 8\pi\rho\qquad\qquad\eqno(2.7) 
$$
$$
{G_x}^x\ = \ddot A  + \tx{1\over4}(3\Adot^2 + \psidot^2) = -8\pi p\qquad\qquad
\eqno(2.8)
$$
$$
\eqalignno{{G_y}^y\ &=\ {G_z}^z\cr
&=\ \tx{1\over2}\ddot A + \tx{1\over4}(\Adot^2 
+ \psidot^2) + \tx{1\over4}\Adot\dot\lambda
+ {\tx{1\over 4}}\dot\lambda^2 + {\tx {1\over2}}\ddot\lambda\cr
&=\ -8\pi p&(2.9)\cr}
$$
$$
{G_y}^z\ =\ \ddot\psi + \Adot\psidot + 
\tx{1\over2}\dot\lambda\psidot = 0\qquad\qquad\qquad\qquad\eqno(2.10)
$$

The matter-response equation for the perfect-fluid {\bi T}, (2.5), is
$$
\dot\rho\ = \ -(\rho + p)(\Adot + \tx{1\over2}\dot\lambda)\eqno(2.11)
$$

\section{Vacuum Solution}

With $\rho = p = 0$, it is convenient to form the combination 
${1\over2}({G_t}^t + {G_x}^x + 2{G_y}^y)$, \ie $\ddot\lambda + 
\dot\lambda(\Adot + {1\over2}\dot\lambda) = 0$, which is immediately 
integrated:
$$
(e^{\lambda/2})_{,t} = c/R^2\eqno(3.1)
$$
Likewise, the combination ${1\over2}({G_t}^t + {G_x}^x)$, \ie
$\ddot A + \Adot^2 + {1\over2}\Adot\dot\lambda = 0$, gives
$$
\Adot = {c_1\over{R^2e^{\lambda/2}}}\eqno(3.2)
$$
Combining (3.1) and (3.2)
leads to a second-order equation for $R$ with solution:
$$
R^2 = R^2_0(t/t_0)^{c_1\over{c + c_1}}\eqno(3.3)
$$
Then
$$
e^{\lambda} = (c + c_1)^2(t/t_0)^{2c\over{c+c_1}}\eqno(3.4)
$$

Now, integrating (2.10) and combining the result with (3.3) and (3.4) yields
$$
\psi = \psi_0 + {c_0\over{c+c_1}}\log(t/t_0)\eqno(3.5)
$$
In these equations, $c$, $c_0$, and $c_1$ are constants 
of integration
which are not all independent. Indeed (2.7) and (2.8) (with $\rho=p=0$) 
are satisfied only if $\kappa^2\equiv(c_0/c_1)^2 = 1 + 4c/c_1$. From this,
the final form for the vacuum solution becomes:
$$
\eqalignno{
e^{\lambda - \lambda_0}\ &=\   (t/t_0)^{{-2(1-\kappa^2)}
\over{3 + \kappa^2}}\cr
R/R_0\ &=\ (t/t_0)^{2\over{3+\kappa^2}}\cr
\psi\ &= \psi_0 + {{4\kappa}\over{3+\kappa^2}}\log(t/t_0)
&(3.6)\cr}
$$
Alternatively, a compact form for $\beta$ and $f$ is written in terms
of the complex parameter $e^q \equiv \beta + \i f = R^2e^{\i\psi}$:
$$
e^{q-q_0} = (t/t_0)^{{4(1+\i\kappa)}\over{3 + \kappa^2}}\eqno(3.7)
$$
As in Einstein's General Relativity, whose corresponding solution 
(Kasner 1921) can be
recovered by setting $\kappa = 0$ and $\psi_0 = n\pi$, $n = 0, 2, 4...$,
the solution is open in the plane of symmetry and closed in the perpendicular
direction if $\kappa^2 < 1$. If $\kappa^2 > 1$, however, the solution is open
in all three spatial directions.

Assuming that the metric is given by ${\sl g}_{\sym{\mu\nu}}$, its
signature changes sign every time $\cos\psi = 0$ since $\beta = R^2\cos\psi$.
This occurs an infinite number of times over a finite interval starting
at $t=0$, although the interval can be made as small as desired by
decreasing the NGT parameter $\kappa$.

At $t=0$, both $R$ and the 4-volume element $\sqrt{-{\sl g}}$ go to
zero, whereas ${\sl g}_{11}\ (-e^\lambda)$ goes to infinity if
$\kappa^2 < 1$ and to zero if $\kappa^2 > 1$. The solution would
thus appear to contain an essential singularity. This is supported
by a computer calculation of the Kretschmann scalar, $R_{\alpha\beta\mu\nu}
R^{\alpha\beta\mu\nu}$, with a symbolic manipulator; the result,
which we do not reproduce here, contains an additive term that goes
like $1/t^4$, along with other terms, also in $1/t^4$, but whose 
numerator oscillates infinitely fast at $t=0$.

\section{Solution with Matter}

In this case, it is best to start from a combination which contains neither
$\rho$ nor $p$. If we subtract (2.9) from (2.8), we obtain after some
rearrangement an equation which is also free of $\psi$:
$$
(\log\,R^2e^{-\lambda})_{,tt} + (\log\,R^2e^{\lambda/2})_{,t}
(\log\,R^2e^{-\lambda})_{,t} = 0\eqno(4.1)
$$
At this point, it is convenient to change the independent variable $t$ to
$$
v\equiv \int{dt\over R^2e^{-\lambda/2}}\eqno(4.2)
$$
Then we obtain from (4.1)
$$
(\log\,R^2e^{-\lambda})_{,vv} = 0\eqno(4.3)
$$
whose solution reads
$$
R^2 = e^{\lambda + \alpha(v - v_0)}\eqno(4.4)
$$
whereas (2.10) becomes $\psi_{,vv} = 0$ and has for solution:
$$
\psi =  \sigma(v - v_0)\eqno(4.5)
$$
In (4.4) and (4.5), $\alpha$ and $\sigma$ are arbitrary constants.

Now we assume an adiabatic equation of state for the matter: $p = \gamma\rho$,
with $0\le\gamma\le1$. From (2.7) and (2.8), the combination
${1\over2}({G_t}^t + {G_x}^x)$, expressed in terms of $v$, is
$$
A_{,vv} = 8\pi(1 - \gamma) R^4e^{\lambda}\rho\eqno(4.6)
$$
With all these results, we can write (2.7) in terms of $v$ and of a
new dimensionless dependent variable $x$:
$$
x_{,v} = \sqrt{6\pi b}(1 - \gamma)\left(x^2 - \eta^2\right)
^{\gamma\over{\gamma - 1}}\eqno(4.7)
$$
where 
$$
x^2\equiv R^4e^{\lambda}\rho/b + \eta^2\eqno(4.8)
$$
and $\eta^2\equiv (\sigma^2 + \alpha^2/3)/32\pi b$, with $b$ an arbitrary
constant. The integral of (4.7) is
$$
x = -\eta\,\coth[\sqrt{6\pi b}\,\eta(1-\gamma)(v - v_0)]
\eqno(4.9)
$$
Now, from the conservation equation (2.11), $\rho(R^2e^{\lambda/2})
^{1 + \gamma}$ is equal to a constant which we identify with $b$, and this
allows us to write (4.8) as $x^2 - \eta^2 = (R^2e^{\lambda/2})^{1 - \gamma}$.

Then, with the help of (4.4), we can  eliminate $\exp[3\alpha(1 - \gamma)v/2]$
between (4.8) and (4.9) to find  an expression for $R$ as
a function of $x$:
$$
R = \left\lbrack(x + \eta)^{1 - \alpha/\sqrt{\alpha^2 + 3\sigma^2}}
(x - \eta)^{1 + \alpha/\sqrt{\alpha^2 + 3\sigma^2}}
\right\rbrack^{1\over{3(1-\gamma)}}\eqno(4.10)
$$
Also, by virtue of (4.9), (4.5) becomes
$$
\psi =  {4\over\sqrt{3}}{\sigma\over\sqrt{\alpha^2/3 +\sigma^2}}
{1\over{\gamma - 1}}{\hbox{arctanh}}\,\left({\eta\over x}\right)\eqno(4.11)
$$
Finally, from (4.4), there comes
$$
e^{\lambda} = \left\lbrack(x+\eta)^{1 - 2\alpha/\sqrt{\alpha^2 + 3\sigma^2}}
(x - \eta)^{1 + 2\alpha/\sqrt{\alpha^2 + 3\sigma^2}}
\right\rbrack^{2\over{3(1-\gamma)}}
\eqno(4.12)
$$
To find $x$ as a function of time, we differentiate (4.8) with respect
to time and note that $\dot v = (R^2e^{\lambda/2})^{-1}$. Then
$$
\sqrt{6\pi b}(1 - \gamma)(t - t_0)= \int(x^2 - \eta^2)^{\gamma\over
{1 - \gamma}}\,dx\eqno(4.13)
$$

Equations (4.10)--(4.13) constitute the general time-dependent
plane-symmetric NGT solution
for an adiabatic equation of state and ${\sl g}_{\asym{0i}} = 0$ 
It reduces to the solution obtained in General Relativity 
(Sch\"ucking and Heckmann 1958) in the limit $\sigma\rightarrow0$.
From now on, however, we put equal to zero the constant $\alpha$  
characterising the anisotropy due to matter in (4.4); any remaining
anisotropy is due to NGT. In that case we are left with
$$
\eqalignno{
R\ &=\ \left(x^2 - \eta^2\right)^{1\over{{3(1-\gamma)}}}\cr
\psi\ &=\ {4\over\sqrt{3}}{1\over{\gamma - 1}}{\hbox{arctanh}}
\left({\eta\over x}\right)\cr
e^{\lambda}\ &=\ R^2\cr
\rho\ &= \rho_0(R/R_0)^{-3(1+\gamma)}
&(4.14)\cr}
$$
where now, $\eta = \sigma/\sqrt{32\pi b}$, a dimensionless constant 
independent of initial
conditions. We can also write the first two equations of (4.14) as
$$
\beta + \i f = \left\lbrack(x + \eta)^{1 - \i\sqrt{3}}(x - \eta)^{1 + 
\i \sqrt{3}}\right\rbrack^{2\over{3(1 - \gamma)}}\eqno(4.15)
$$
The dependence of $x$ on time is to be found from (4.13). For instance,
if $\gamma=0$ (matter-dominated universe), $6\pi b\,t^2 = x^2$, so that
$$
R = R_0(6\pi\rho_0t^2 - \eta^2/R_0^3)
^{1/3}\eqno(4.16)
$$
whereas, if $\gamma= 1/3$ (radiation-dominated universe), $R^2 =
(x^2 - \eta^2)$, with $x$ given by the implicit relation
$$
\sqrt{6\pi\rho_0}\,t = {3\over{4R_0}}\left\lbrack
x(x^2 - \eta^2)^{1/2} - \eta^2\log\left(x + 
(x^2 - \eta^2)^{1/2}\right)\right\rbrack\eqno(4.17)
$$
It is clear from (4.14) and (4.15) that this solution becomes isotropic
in the large $x$ (large time) limit, in the sense that $f\rightarrow0$ and
$\beta\rightarrow e^{\lambda}$, so that we recover the 
Friedmann-Robertson-Walker solution of General Relativity. It is equally
clear that at some time $t_{\hbox{s}} < t_0$, there occurs
a branch-point at $x = \eta$; $R$ and $e^\lambda$ go to zero, 
and the matter density becomes infinite. When $\gamma = 0$, 
$$
\eqalignno{
t_{\hbox{s}}\ &= \eta/\sqrt{6\pi b} \cr
                   &= \ {{\sigma/\sqrt{3}}\over{8\pi b}}&(4.18)\cr}
$$
One can show from (4.13) that, both for a dust and
a radiation universe, the interval between $t_0$ and $t_{\hbox{s}}$
is smaller in the NGT solution than in the corresponding one in GR.

\section{Early-time Behaviour and Correction to the Age of the Universe}

At $t = t_0$, the angle $\psi$ takes the value 
$$
\psi_0 = {4\over\sqrt{3}}{1\over{\gamma-1}}{\hbox{arctanh}}
\sqrt{\eta^2\over{R_0^3+\eta^2}}\eqno(5.1)
$$
whereas $\psi(t_{\hbox{s}})\rightarrow-\infty$. Thus, $\beta$ and $f$ must
change signs an infinite number of times between $t_{\hbox{s}}$ and
$t_0$. The zeros of $\beta$ occur when $\cos\,\psi = 0$, at
$$
x_n = \eta\coth\left\lbrack{\sqrt{3}\over8}\pi(1 - \gamma)(2n - 1)\right\rbrack
\qquad\qquad (n = 1, 2,\ldots)\eqno(5.2)
$$
For a dust-filled universe, the first zero of $\beta$ is at
$x_1 = 1.7\eta\approx \sigma/6b$. All other zeros lie between $\eta$ and
$1.7\eta$. The time elapsed between $t_1$ and $t_0$ is
$$
t_0 - t_1 = {1\over\sqrt{6\pi\rho_0}}\left\lbrack
\left( 1 + {\eta^2\over R_0^3}\right)^{1/2} - 1.7{\eta\over R_0^{3/2}}
\right\rbrack\eqno(5.3)
$$
The behaviour of $\beta$ over the interval $0<t<t_1$ has no obvious
physical meaning, but it can be shrunk to unobservable size by reducing
the value of $\eta$ (or $\sigma$). Then, if we take $t_0$ as the present time, 
expression (5.3) can be interpreted as a rough estimate of the age of 
the universe. 

Now, another way of expressing the age of the universe is with the Hubble
parameter, here defined as the average rate of expansion of its
3-volume:
\hbox{$H = {1\over3}(\log\beta e^{\lambda/2})_{,t}$},
or, if we use the ($R,\,\psi$) representation, 
$H = \dot R/R - {1\over3}\psidot\tan\,\psi$. Also, (2.8) can be 
cast in the form
$$
\left({\dot R\over R}\right)^2 = {{8\pi}\over3}\rho + {1\over12}\psidot^2
\eqno(5.4)
$$
Insertion of the solution with matter then yields, to first order in
$(\sigma/R_0^3)^2$:
$$
\eqalignno{
H^2 &= {{8\pi}\over3}\rho + {1\over12}\left({\sigma\over R^3_0}\right)^2
\left({\rho\over\rho_0}\right)^{2\over{1+\gamma}}
\left(1 + {16\over{3(1 -\gamma)}}\right)\cr
&={{8\pi}\over3}\left(R\over R_0\right)^{-3(1-\gamma)} 
+ {1\over12}\left({\sigma\over R^3_0}\right)^2
\left({R\over R_0}\right)^{-6}
\left(1 + {16\over{3(1 -\gamma)}}\right)&(5.5)\cr}
$$
From this comes a relation, valid at the present time,  between the 
Hubble parameter, the matter density, and $\sigma/R_0^3$:
$$
H^2_0 = {{8\pi}\over3}\rho_0 + {19\over36}\left(\sigma\over R_0^3\right)^2
\eqno(5.6)
$$
Now we can express our ``age of the universe'' in (5.3) as
$$
t_0 - t_1 \approx {2\over3H_0}\left(1 - {1\over15}
{\sigma\over b}
\right)\eqno(5.7)
$$
This last form is convenient because both $\sigma$ and $b = \rho R^3$ remain
constant throughout the time evolution. But no very stringent bound
can be inferred from it. For instance, if $\sigma/b = 0.1$, the length
of the matter-dominated era is shortened by one percent only.
 
In a radiation-dominated universe ($\gamma= 1/3$), we have instead
$$
H^2 = {{8\pi}\over3}\rho  + {3\over4}\left({\sigma\over 
R_0^3}\right)^2\left({\rho\over\rho_0}\right)^{3/2}\eqno(5.8)
$$
or, in terms of black-body radiation temperature,
$$
H^2 = {{8\pi a}\over3}T^4\left\lbrack 1 + {9\over{32\pi a}}\left({\sigma\over 
T_0^2 R_0^3}\right)^2\left({T\over T_0}\right)^2
\right\rbrack\eqno(5.9)
$$
where $a$ is the Stefan constant, and $\sigma$ is the same as in
the dust solution so as to ensure proper matching of the two cases.

\section{Birefringent Universe}

As a nonmetric theory of gravitation, NGT couples matter fields directly
to ${\sl g}_{\asym{\mu\nu}}$ in a way that generically violates the 
Einstein equivalence principle. For instance, the propagation of
electromagnetic waves should be affected by the coupling between
${\sl g}_{\asym{\mu\nu}}$ and electromagnetic fields. This was first 
pointed out by Mann and Moffat (1981) and detailed calculations were 
carried out (Will 1989, Gabriel \etal 1990, 1991a, 1991b, 1991c) for 
${\sl g}_{\asym{i0}}\neq 0$, ${\sl g}_{\asym{ij}}=0$ 
in the static spherically-symmetric case. In the cosmological 
plane-symmetric solution with matter discussed here, we should also find
some  dependence of the propagation of light on its polarisation.

The coupling of NGT to electromagnetism is written as an
electromagnetic Lagrangian density  which has the general form 
(Mann \etal 1989)
$$
{\cal L}_{\rm em} = - {1\over16\pi}\sqrt{-{\sl g}}{\cal F}
{\sl g}^{\mu\alpha}{\sl g}^{\nu\beta}\lbrack Z F_{\mu\nu}F_{\alpha
\beta} + (1 - Z)F_{\alpha\nu}F_{\mu\beta} + Y F_{\mu\alpha}
F_{\nu\beta}\rbrack\eqno(6.1)
$$
where $Y$ and $Z$ are constants, and ${\cal F} = 1$ when 
${\sl g}_{\asym{\mu\nu}}= 0$. As usual, the electromagnetic field
tensor, $F_{\mu\nu}$, can be decomposed into electric components,
$F_{i0}= E_i$, and magnetic components, 
$F_{ij}= \epsilon_{ijk} B^k$. We also take ${\cal F} =
\sqrt{-{\sl g}}/\sqrt{-\det\,{\sl g}_{\sym{\mu\nu}}}$.
With (2.1) and $R^2 = e^\lambda$ in (4.14), (6.1) can then be written
$$
\eqalignno{
{\cal L}\ &= {1\over8\pi}\left\lbrack\epspar{\bi E}_\parallel^2 +
\epsperp{\bi E}_\perp^2 - \left({{\bi B}_\parallel^2\over\mupar} +
{{\bi B}_\perp^2\over\muperp}\right)\right\rbrack\cr
&= {1\over8\pi}\left\lbrack\epspar\left( {\bi E}^2 + 
({\epsperp\over\epspar} - 1)({\bi n} \cdot{\bi E})^2\right) - 
{1\over\mupar}\left({\bi B}^2 - \left(1 - {\mupar\over\muperp}\right)
({\bi n} \cdot{\bi B})^2\right)\right\rbrack&(6.2)\cr}
$$
where 
$$
\eqalignno{
\epspar\ &= \mupar = R\cr
\epsperp\ &\equiv \epspar/\cos\,\psi\cr
\muperp\ &= \mupar\cos\,\psi(1 - 2X\sin^2\psi)^{-1}&(6.3)\cr}
$$
and ${\bi n}$ is a unit vector perpendicular to the plane of symmetry.
One may think of $\bgr\epsilon$ as a diagonal tensor with 
nonzero components
$(\epsilon_{xx},\epsilon_{yy},\epsilon_{zz})=(\epsperp,\epspar,\epspar)$,
and similarly for a diagonal tensor $\bgr\mu$. Our choice of ${\cal F}$
ensures that $\epspar = \mupar$, so that any NGT 
effect must arise from the
$({\bi n} \cdot{\bi E})^2$ and $({\bi n} \cdot{\bi B})^2$ terms in 
${\cal L}$.

The Maxwell-NGT equations in momentum space which follow from varying (6.2) are
$$
\eqalignno{
{\bi k}\cdot{\bi D}\ &= \ 0\cr
{\bi k}\cdot{\bi B}\ &= \ 0\cr
{\bi k}\wedge{\bi E} - \omega {\bi B}\ &= \ 0\cr
{\bi k}\wedge{\bi H} + \omega {\bi D}\ &= \ 0&(6.4)\cr}
$$
where we have defined ${\bi D}\equiv\bgr{\epsilon}\cdot{\bi E}$,
and ${\bi B}\equiv \bgr{\mu}\cdot{\bi H}$. Note that
${\bi E}$ and ${\bi H}$ are not perpendicular to ${\bi k}$, 
but ${\bi D}$ and ${\bi B}$ are. To derive a
dispersion relation for electromagnetic waves, it is reasonable to
treat $\bgr\epsilon$ and $\bgr\mu$ as constants over
one wavelength. One then obtains
$$
(\epspar\mupar\omega^2 - k^2){\bi B} + \left(1 - {\mupar\over\muperp}\right)
[k^2{\bi n} - {\bi k}({\bi n}\cdot{\bi k})]({\bi n}\cdot{\bi B}) +
\left(1 - {\epspar\over\epsperp}\right)[{\bi B}\cdot{\bi n}\wedge{\bi k}]
({\bi n}\wedge{\bi k}) = 0\eqno(6.5)
$$

This leads to two possible speeds for the waves, one for waves with 
polarisation perpendicular to the direction of anisotropy, ${\bi n}$, 
and one for waves polarised in the plane defined by ${\bi n}$ and ${\bi k}$. 
In the first case, ${\bi n}\cdot{\bi B} = 0 $, and the speed is
$$
c_{\perp} = {1\over R}\left(1 - (1 -\cos\,\psi)\sin^2\theta\right)^{1/2}
\eqno(6.6)
$$
where $\theta$ is the angle between the direction of anisotropy
and the direction of propagation of the waves.

On the other hand, when their magnetic field lies in the plane that
contains ${\bi n}$ and ${\bi k}$, waves propagate at speed
$$
c_{\parallel} = {1\over R}\Big(1 -\left\lbrack(1 - \cos\,\psi) + (2X -1)
\sin^2\psi/\cos\,\psi\right\rbrack\sin^2\theta\Big)^{1/2}\eqno(6.7)
$$

There are two contributions to (6.6) and (6.7) which should be carefully
distinguished. One arises simply from the fact that the waves
propagate on a NGT plane-symmetric background. Indeed, consider a
source situated at a distance $r$ from the Earth, with the vector $\bhat r$ 
making an angle $\theta$ with ${\bi n}$ (or $\bhat x$).
From (2.1), the equation for light propagation is
$$
dt^2 - e^{\lambda} dx^2 - \beta(dy^2 + dz^2) = 0\eqno(6.8)
$$
Then, in the ($R,\psi$) representation, and using $e^\lambda = R^2$
from (4.14), the speed of the light emitted by the source
as measured by an observer on Earth would be
$$
{{dr}\over{dt}} = {1\over R}\left\lbrack1 - (1
-\cos\,\psi)\sin^2\theta\right\rbrack^{1/2}
\eqno(6.9)
$$
which is precisely $c_{\perp}$. In fact, only the term proportional
to $\sin^2\psi$ in $c_{\parallel}$ is a consequence of the direct
coupling between NGT and electromagnetism.

The resulting effect bears a lot of similarity to birefringence
in crystals with one optical axis. The polarisation vector of an 
electromagnetic wave rotates along its trajectory in a way that depends
on the difference between the two speeds. Assuming that $\psi$ is very
small all along the trajectory in a dust-filled universe ($\gamma = 0$), 
this is:
$$
\eqalignno{
c_{\parallel} - c_{\perp}\ &\approx {1\over R}(X- {\tx{1\over2}})
\sin^2\psi\sin^2\theta\cr
&\approx {1\over R}(X- {\tx{1\over2}})\left({\sigma\over R_0^3}\right)^2
{1\over{(6\pi\rho_0)^2}}{1\over t^2}&(6.10)\cr}
$$
The only parameter-free prediction here is that waves propagating
along the anisotropic direction will experience no rotation of their
polarisation vector.

\section{Conclusion}

Equations (3.6) give the solution to the vacuum homogeneous,
plane-symmetric case in NGT, whereas equations (4.14) represent
the equivalent solution in the presence of a perfect fluid.
Both exhibit zeros in ${\sl g}_{22}$ and in ${\sl g}_{\asym{23}}$.
The actual physical meaning of this behaviour is not known
at present. In the solution with matter, the NGT parameter $\sigma$
can be chosen so that the zeros occur 
early enough in the evolution of the universe for them to be
essentially indistinguishable from the initial singularity.
Thereafter this anisotropic universe approaches 
a FRW universe in such a way that currently well-established features
of its early history (such as helium production) do not
change significantly. 

It is interesting that these signature reversals do not
occur in a complex Hermitian version of NGT, with
${\sl g}_{\mu\nu} = {\sl g}_{\sym{\mu\nu}} + e\,{\sl g}_{\asym{\mu\nu}}$,
where $e^2 = -1$. Taking $\beta = R^2\cosh\psi$ and
$f = R^2\sinh\psi$, the solution (3.6) now becomes
$$
\eqalignno{
e^{\lambda - \lambda_0}\ &=\   (t/t_0)^{{-2(1+\kappa^2)}
\over{3 - \kappa^2}}\cr
R/R_0\ &=\ (t/t_0)^{2\over{3-\kappa^2}}\cr
\psi\ &= \psi_0 + {{4\kappa}\over{3-\kappa^2}}\log(t/t_0)
&(7.1)\cr}
$$
with $\kappa$ a real constant. Therefore, $\beta$ goes like
$t^{{4(1+\kappa)}\over{3-\kappa^2}}(1 + t^{-{8\kappa}\over
{3-\kappa^2}})$ and does not go through zero. If $\kappa^2>1$,
it even has a minimum at $t = ({\kappa -1\over{\kappa + 1}})^
{{3-\kappa^2}\over{8\kappa}}$.

In the case with matter, the solution (4.14) becomes
$$
\eqalignno{
R\ &=\ \left(x^2 + \eta^2\right)^{1\over{{3(1-\gamma)}}}\cr
\psi\ &=\ {4\over\sqrt{3}}{1\over{\gamma - 1}}{\hbox{arctan}}
\left({\eta\over x}\right)\cr
e^{\lambda}\ &=\ R^2\cr
\rho\ &= \rho_0(R/R_0)^{-3(1+\gamma)}
&(7.2)\cr}
$$
where $\eta = \sigma/\sqrt{32\pi b}$ is real, and $x$ is still
obtained from (4.13). As $t\rightarrow0$ (or $x\rightarrow0$),
$\psi\rightarrow (2\pi/\sqrt{3})/(\gamma -1)$, and 
$\beta\rightarrow\eta^{4/3(1-\gamma)}\cosh[(2\pi/\sqrt{3})/(1 -\gamma)]$,
again without any oscillation through zero.

It has been asserted (Kelly and Mann 1986, Mann and Moffat 1982)
that only the hypercomplex structure leads to a linearised NGT that is free
of ghost poles. Yet, as mentioned in the introduction, hypercomplex NGT
does have ghost poles generically if the ${\sl g}_{\asym{0i}}$ modes are
not somehow decoupled to all orders.
Thus, it may be premature to reject the Hermitian structure, especially since,
at least in the cases discussed here, the solutions do
not exhibit the puzzling signature reversals of the hypercomplex theory.
Whether their 
singularity structure differs from that encountered in GR, as suggested
by (7.1) and (7.2), can only be established by a more thorough investigation.

Finally, an intriguing possibility is that some small residual anisotropy
may affect the propagation of light in such a way that its
polarisation vector rotates as it travels over cosmological
distances. Such an effect has been reported recently
(Nodland and Ralston 1997a, 1997b, 1997c), but the validity
of these observations
has been disputed (Carroll and Field 1997, Eisenstein and
Dunn 1997, Leahy 1997, Loredo \etal 1997, Wardle
\etal 1997). If they were confirmed, NGT would provide a possible
explanation; if not, an upper bound could be obtained on the size of the
coupling of NGT to electromagnetism.

\ack
The author is grateful to the Instituut voor Theoretische Fysika of
the Katholieke Universiteit in Leuven, Belgium, where most of this work
was done, for their kind hospitality. He also thanks J. W. Moffat
for some helpful comments.

\references

\refjl{Burko L and Ori A 1995}{\PRL}{75}{2455}
\refbk{Carroll S M and Field G B 1997}{}{}{\tt astro-ph/9704263}
\refjl{Clayton M A 1996}{\CQG}{13}{2851}
\refjl{\dash 1997}{\sl Int. Journ. Mod. Phys. {\rm\ A}}{12}{2437}
\refjl{Cornish N J and Moffat J W 1994}{\PL}{B336}{337}
\refjl{Crumeyrolle A 1967}{\sl Riv. Mat. Univ. Parma}{8}{27}
\refjl{Einstein A 1945}{\sl Ann. Math.}{46}{578}
\refjl{Einstein A and Straus E G 1946}{\sl Ann. Math.}{47}{731}
\refbk{Eisenstein D J and Bunn E F 1997}{}{}{\tt astro-ph/9704247}
\refjl{Gabriel M D, Haugan M P, Mann R B, and Palmer J H 1991a}
{\PR {\rm\ D}}{43}{308}
\refjl{\dash 1991b}{\PR {\rm\ D}}{43}{2465}
\refjl{\dash 1991c}{\PRL}{67}{2123}
\refjl{Kasner E 1921}{\sl Am. J. Math.}{43}{217}
\refjl{Kelly P F and Mann R B 1986}{\CQG}{3}{705}
\refjl{Kunstatter G, Moffat J W and Malzan J 1983}{\JMP}{24}{886}
\refbk{Leahy J P 1997}{}{}{\tt astro-ph/9704285}
\refjl{L\'egar\'e J and Moffat J W 1995}{\sl Gen. Rel. Grav.}{7}{761}
\refbk{Loredo T J, Flanagan E E, and Wasserman I M 1997}{}
{\tt astro-ph/9706258}

\refjl{Mann R B 1984}{\CQG}{1}{561}
\refjl{\dash 1989}{\CQG}{6}{41}
\refjl{Mann R B and Moffat J W 1981}{\CJP}{59}{1730}
\refjl{\dash 1982}{\PR {\rm\ D}}{26}{1858}
\refjl{Mann R B, Palmer J H, and Moffat J W 1989}{\PRL}{62}{2765}
\refjl{Moffat J W 1979}{\PR {\rm\ D}}{19}{3554}
\refbk{\dash 1991}{Gravitation 1990: A Banff Summer Institute}
{ed R B Mann and P Wesson, (Singapore: World Scientific)}

\refjl{\dash 1984}{\sl Found. Phys.}{14}{217}
\refjl{\dash 1995a}{\PL}{B355}{447}
\refjl{\dash 1995b}{\JMP}{36}{3722}
\refjl{\dash 1995c}{\JMP}{36}{5897}
\refbk{Moffat J W and Sokolov I Yu 1995}{}{\tt UTPT--95--21
\tt astro-ph/9510068}
\refjl{Nodland B and Ralston J P 1997a}{\PRL}{78}{3043}
\refbk{\dash 1997b}{}{\tt astro-ph/9705190}
\refbk{\dash 1997c}{}{\tt astro-ph/9706126}
\refbk{Sch\" ucking E and Heckmann O 1958}{Onzi\`eme Conseil de
Physique Solvay}{(Brussels: Stoops), p 149}
\refjl{Vincent D 1985}{\CQG}{2}{409}
\refbk{Wardle J F C, Perley R A, and Cohen M H 1997}{}
{\tt astro-ph/9705142}

\refjl{Will C M 1989}{\PR {\rm\ D}}{62}{369}
\refjl{Wyman M 1950}{\sl Can. J. Math.}{2}{427} 

\bye